\documentclass[12pt,manuscript]{aastex}

\newcommand{\2}{{~\sc ii}}
\newcommand{\3}{{~\sc iii}}
\newcommand{\4}{{~\sc iv}}

\usepackage{eucal} 

\shorttitle{PAH Strength and the Interstellar Radiation Field around the Massive Young Cluster NGC\,3603}
\shortauthors{Lebouteiller et al.}

\begin{document}


\title{PAH Strength and the Interstellar Radiation Field around the Massive Young Cluster NGC\,3603}


\author{V. Lebouteiller}
\affil{Center for Radiophysics and Space Research, Cornell University, Space Sciences Building, Ithaca, NY 14853-6801, USA}
\email{vianney@isc.astro.cornell.edu}

\author{B. Brandl}
\affil{Leiden Observatory, University of Leiden, P.O. Box 9513, 2300 RA Leiden, Netherlands}
%\email{brandl@strw.leidenuniv.nl}

\and

\author{J. Bernard-Salas, D. Devost, and J.~R. Houck}
\affil{Center for Radiophysics and Space Research, Cornell University, Space Sciences Building, Ithaca, NY 14853-6801, USA}

%\altaffiltext{1}{Visiting Astronomer, Cerro Tololo Inter-American Observatory.
%CTIO is operated by AURA, Inc.\ under contract to the National Science


\begin{abstract}
We present spatial distribution of polycyclic aromatic hydrocarbons and ionized gas 
within the Galactic giant H\2\ region NGC\,3603. Using the \textit{IRS}
instrument on board the Spitzer Space Telescope, we study in particular the PAH emission
features at $\sim$5.7, 6.2, 7.7, 8.6, and 11.3\,$\mu$m, and the [Ar\2] 6.99\,$\mu$m,
[Ne\2] 12.81\,$\mu$m, [Ar\3] 8.99\,$\mu$m, and [S\4] 10.51\,$\mu$m forbidden emission
lines. The observations probe both ionized regions and photodissociation regions.
Silicate emission is detected close to the central cluster while silicate absorption is
seen further away. We find no significant variation of the PAH ionization fraction across
the whole region. The emission of very small grains lies closer to the central stellar
cluster than emission of PAHs. The PAH/VSG ratio anticorrelates with the hardness of the
interstellar radiation field suggesting a destruction mechanism of the molecules within
the ionized gas, as shown for low-metallicity galaxies by Madden et al.\ (2006).
\end{abstract}



\section{Introduction}

The investigation of star-formation feedback on the interstellar medium (ISM) in low-metallicity environments is of prime
importance to understand and constrain the evolution of primordial systems.
The most relevant diagnostics are drawn from the study of the various ISM components as a function of the physical
conditions in starburst-dominated objects such as dwarf star-forming galaxies and giant H\2\ regions.
In this perspective, the mid-infared (MIR) domain, which gives access to both gas and dust components,
provides one of the most powerful tools.

Star-forming objects are characterized by strong MIR emission features attributed to polycyclic
aromatic hydrocarbons (PAHs)
in the photodissociation envelopes surrounding massive star clusters (L{\'e}ger \& Puget 1984;
Allamandola et al.\ 1985, 1989; Puget \& L{\'e}ger 1989; Tielens et al.\ 1999).
The PAHs emitting in the $3$-$13$\,$\mu$m range contain from several tens up to several hundreds carbon atoms
(see, e.g., Schutte et al.\ 1993). PAH molecules are mainly excited by far-UV radiation.
The MIR emission features are due to the subsequent fluorescence of aromatic C$-$C (especially
dominant at 6.2 and 7.7\,$\mu$m) and peripheral C$-$H
(3.3, 8.6, and 11.3\,$\mu$m) fundamental vibrational and bending modes.

The various PAH emission features are diffe\-ren\-tly affected by the local physical
conditions such as the hardness of the interstellar radiation field (ISRF), the dust
temperature, or the PAH mixture (Hony et al.\ 2001; Peeters et al.\ 2002). It has been
suggested that strong UV radiation is able to ionize the PAHs, while for weaker
radiation, PAHs can be neutral or even negatively charged by accretion of a single
electron (Bakes \& Tielens 1994, 1998; Salama et al.\ 1996; Dartois \& d'Hendecourt
1997). Laboratory experiments indicate that neutral PAHs show stronger C$-$H mode emission
relatively to C$-$C modes while the inverse is true for ionized PAHs (Szczepanski \&
Vala 1993; Langhoff 1996; Kim et al.\ 2001; Hudgins \& Allamandolla 1999). As a result,
the emission features at 3.3 and 11.3\,$\mu$m are thought to mainly originate from
neutral PAHs while the emission features between 6 and 9\,$\mu$m are due to ionized PAHs.
Recently, it has been demonstrated that not only the UV radiation but also the
metallicity and the dust extinction are able to significantly influence the PAH
ionization fraction (Cox~\&~Spaans 2006).

Because of their chemical composition, PAH molecules are expected to be less abundant in low-metallicity environments.
The PAH intensity in galaxies has been found to correlate with the ISM metallicity (see e.g., Madden et al.\ 2006;
Engelbracht et al.\ 2005; Wu et al.\ 2006).
This correlation can be due either to the low carbon abundance, to the harder ISRF from
low-metallicity hot stellar atmospheres (see e.g., Schaller et al.\ 1992;
Schaerer \& Maeder 1992;Schaerer et al.\ 1993; Charbonnel et al.\ 1993), or to a combination of these two effects.
Madden et al.\ (2006) showed that the radiation field itself has an impact on the PAH survival
in various metallicity dwarf galaxies.

The PAH spectrum is seen to vary not only from one object to another but also within a single
object. Hence it is possible to investigate the variations of the PAH spectrum as a
function of the physical conditions across a given region (see e.g., Joblin et al.\ 2000;
Vermeij et al.\ 2002; Bregman \& Temi 2005; Kassis et al.\ 2006). In a few objects, the
PAH intensity has been found to decrease when the ISRF hardens (Verstraete et al.\
1996; Madden et al.\ 2006; Beir{\~a}o et al.\ 2006), suggesting that PAH molecules are
destroyed by high-energy photons.

The \textit{Infrared Spectrograph} (\textit{IRS}; Houck et al.\ 2004) on board the Spitzer Space Telescope
(Werner et al.\ 2004a) opened a new perspective in the extraction of
small-scale regions within extended source. The wavelength range covered by the \textit{IRS} gives the possibility to
investigate the spatial distribution of the ionized gas, the molecular hydrogen gas, the PAHs, the silicate dust, and
the very small grains (VSGs). One of the most interesting application enabled by the \textit{IRS} is to understand the
influence of the local physical conditions, such as the hardness of the ISRF, on the molecular content.

Galactic giant H\2\ regions are ideal benchmarks for such a study. NGC\,3603 is a giant
H\2\ region located $\sim$7\,kpc from the Sun along the Galactic plane (Moffat 1983;
Moffat et al.\ 1994; Drissen et al.\ 1995; Brandl et al.\ 1999). Oxygen abundance measurements
range from $12 + \log (\mathrm{O}/\mathrm{H}) = 8.39$ to $8.52$ (Melnick et al.\ 1989; Tapia et al.\ 2001; Garc{\'{\i}}a-Rojas et al.\ 2006) and imply
a metallicity close to solar. More than 50
O and WR stars (Moffat et al.\ 1994) produce a Lyman continuum flux of
$10^{51}$\,s$^{-1}$ (Kennicutt 1984; Drissen et al.\ 1995), which is about 100 times the
ionizing power of the Orion Trapezium cluster. Within the core of the cluster, the system
HD\,97950 contains several WR, O3, and many late O stars (Melnick et al.\ 1989; Hofmann
et al.\ 1995). The massive stars influence the surrounding ISM morphology, notably by
compressing the molecular clouds through stellar winds (N{\"u}rnberger \& Stanke 2002). For
this reason, the geometry consists in a complex arrangement of numerous, localized H\2\
region $-$ PDR transitions. Although its bolometric luminosity of $L_{\rm{bol}} \sim
10^7$\,L$_{\odot}$ is only about one tenth of the luminosity of 30\,Doradus, it looks
remarkably similar to R136, the stellar core of 30\,Doradus (Moffat et al.\ 1994). 
NGC\,3603 has often been referenced as the most massive, optically visible H\2\ region in
our Galaxy, due to its high luminosity and moderately low extinction of $A_{\rm{v}} \sim
4.5$ (Eisenhauer et al.\ 1998), most of which is Galactic foreground extinction. 

In this paper, we investigate with the \textit{IRS} the spatial variations of the MIR features across NGC\,3603. 
After introducing the observations in \S\ref{sec:observations}, 
we describe the data analysis in \S\ref{sec:method}.
The MIR morphology is then investigated in \S\ref{sec:irac}.
The gas distribution and in particular the ISRF hardness is derived in \S\ref{sec:isrf}.
Distribution of dust and molecular features is discussed in \S\ref{sec:solid}. 
Finally, we question the PAH survival in \S\ref{sec:pahsurvival}, and apply our results to usual MIR diagnostic 
diagrams in \S\ref{sec:diagnostic}.

\section{Observations}\label{sec:observations}

\begin{figure*}
\includegraphics[angle=0,scale=.51,clip=true]{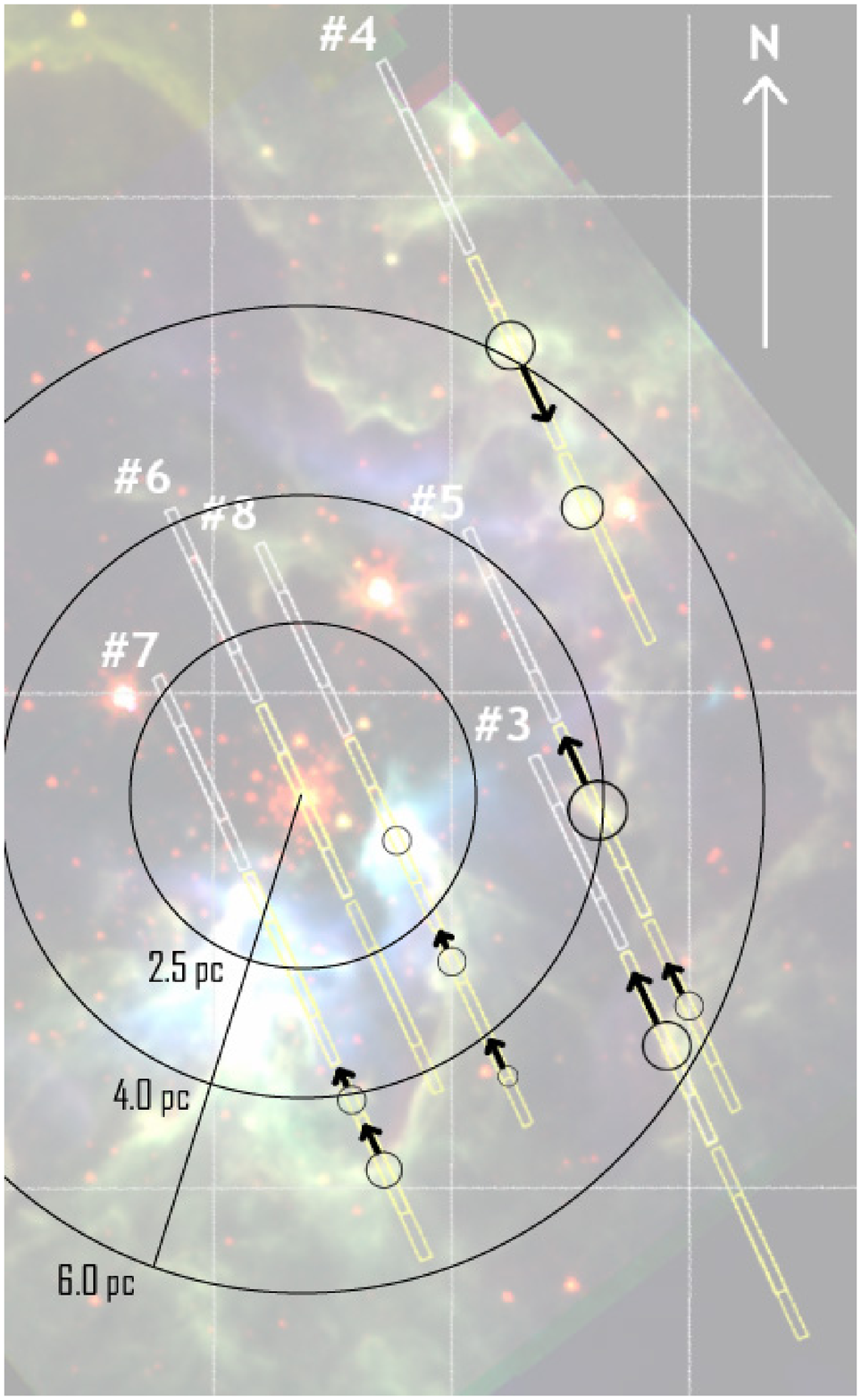}\vspace{0cm}
\hspace{0cm}
\includegraphics[angle=0,scale=.51,clip=true]{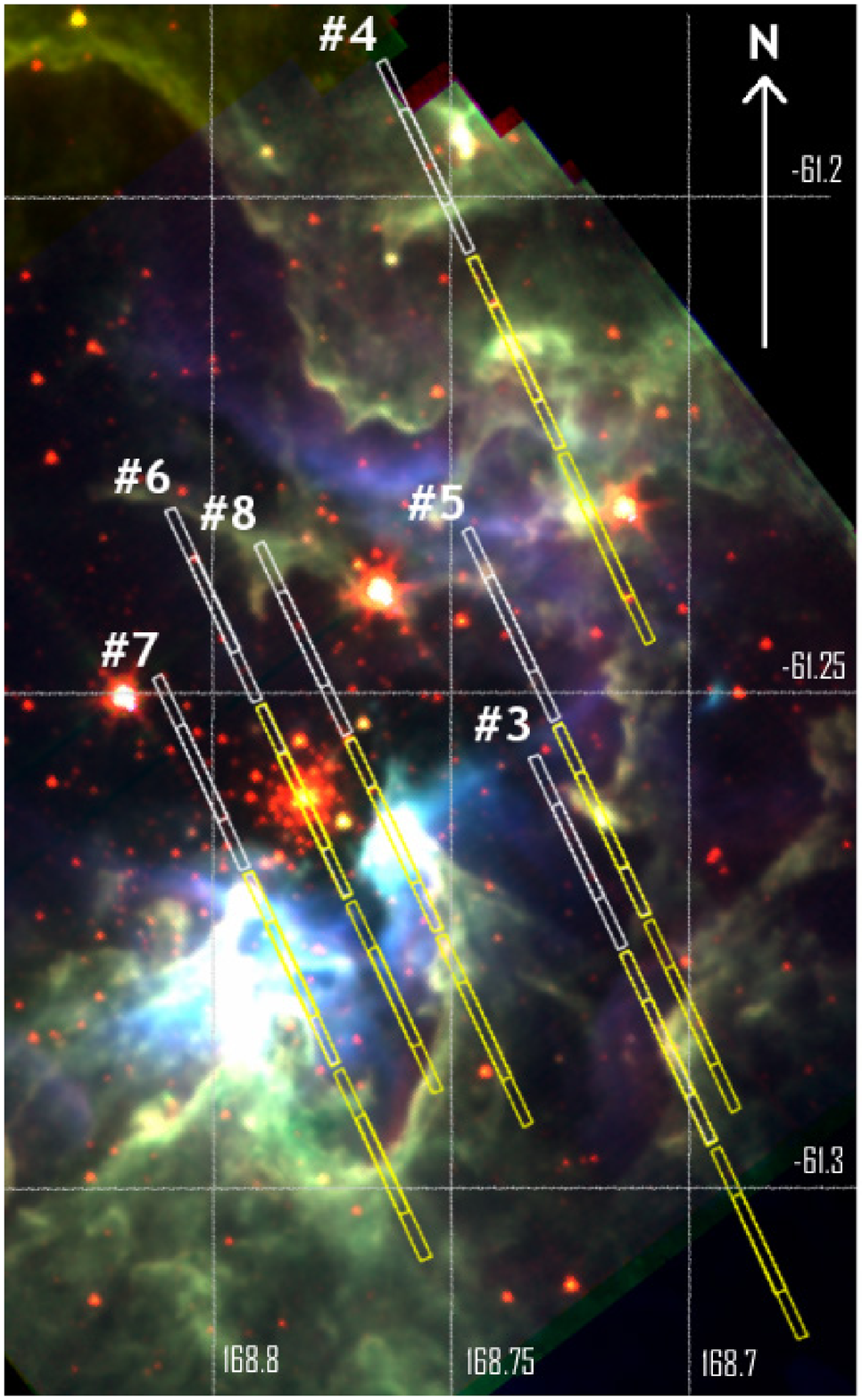}\vspace{0cm}
\figcaption{Slit positions are superimposed on an \textit{IRAC} composite image. The colors
are defined by the \textit{ch1} band ($[$3.17-3.93$]$\,$\mu$m) for the red, \textit{ch3}
($[$5.02-6.44$]$\,$\mu$m) for the green, and \textit{ch4} ($[$6.42-9.32$]$\,$\mu$m) for
the blue. For a given position, the northern slits observe only the second spectral order
(SL2), the southern slits observe only the first spectral order (SL1), and the middle slit
observations give access to both orders.
The local PAH emission maxima (small circles) are plotted on the left panel. The size of the circles
is proportional to the PAH peak intensity (sum of all bands). We also show the regions (arrows) showing an anticorrelation between
PAH intensity and both [Ar\3] and [S\4] line intensities.
\label{fig:slits}}
\end{figure*}

NGC\,3603 was observed as part of the \textit{IRS} GTO program to study massive star
formation in Local Group giant H\2\ regions (PID\,63). Extensive imaging using the
\textit{Infrared Array Camera} (\textit{IRAC}; Fazio et al.\ 2004) was also peformed
(Brandl et al.\ 2007, in preparation), and a total of nine infrared bright sources (clusters, shock
fronts, protostars, etc...) have been selected for follow-up spectroscopy with the high-
and the low-resolution modules of the \textit{IRS} (AORKEY 12080384). The complete spectral analysis will
be discussed in Lebouteiller et al.\ (2007, in preparation). The long slits of the \textit{IRS}
low-resolution modules, however, cover a significant area of NGC\,3603 and provide
information on the spatial variations of spectral features across the region. 

We present an analysis of the positions \#3 to \#8 from the original follow-up
observations (see the coordinates in Table~\ref{tab:positions} and the image of
Figure\,\ref{fig:slits}). Position \#1 is an offset observation and was used for sky
substraction. Positions \#2 and \#9 (the latter corresponding to the IR luminous source IRS-9)
gave corrupted and saturated data respectively. The position \#6 is centered on the
central stellar cluster while the other positions are centered on bright MIR knots
located from $\sim$1.6 to $\sim$6\,pc away from the cluster (we use hereafter the projected distance from the
central stellar cluster, which represents the smallest possible distance).

\begin{deluxetable}{llll}
\tablecolumns{4}
\tablecaption{Requested coordinates of each position.\label{tab:positions}}
\startdata
\tableline\tableline
 \# & $\alpha$ (J2000) & $\delta$ (J2000) & PA\tablenotemark{a} \\
\tableline
1\tablenotemark{b} &$11^\mathrm{h}14^\mathrm{m}21^\mathrm{s}.80$ & $-61^\circ24'20''.0$ & $+24^\circ.62$ \\
2\tablenotemark{c} &$11^\mathrm{h}15^\mathrm{m}17^\mathrm{s}.00$ & $-61^\circ19'10''.0$ & $+24^\circ.42$ \\
\tableline
3 & $11^\mathrm{h}14^\mathrm{m}49^\mathrm{s}.06$ & $-61^\circ17'09''.1$ & $+24^\circ.55$ \\
4 & $11^\mathrm{h}14^\mathrm{m}56^\mathrm{s}.71$ & $-61^\circ12'56''.6$ & $+24^\circ.52$ \\
5 & $11^\mathrm{h}14^\mathrm{m}52^\mathrm{s}.40$ & $-61^\circ15'46''.3$ & $+24^\circ.54$\\
6 & $11^\mathrm{h}15^\mathrm{m}07^\mathrm{s}.40$ & $-61^\circ15'39''.2$ & $+24^\circ.48$  \\
7 & $11^\mathrm{h}15^\mathrm{m}08^\mathrm{s}.03$ & $-61^\circ16'40''.2$ & $+24^\circ.48$\\
8 & $11^\mathrm{h}15^\mathrm{m}02^\mathrm{s}.88$ & $-61^\circ15'51''.6$ & $+24^\circ.50$\\
\tableline
9\tablenotemark{c} & $11^\mathrm{h}15^\mathrm{m}03^\mathrm{s}.30$ & $-61^\circ21'25''.0$ & $+24^\circ.47$ \\
\tableline
\enddata
\tablenotetext{a}{Position angle.}
\tablenotetext{b}{Position \#1 was used for sky substraction.}
\tablenotetext{c}{Position \#2 and \#9 gave unusable data (see text).}
\end{deluxetable}

The low-resolution spectra ($\lambda/\Delta\lambda \sim 65$-$130$)
from the short-low (SL) module cover the spectral region 5.2-14.5\,$\mu$m.
The SL module contains two slits, one for each spectral order. The SL1 slit corresponds to the
first order (7.4-14.5\,$\mu$m) and has a size of $3.7''\times57''$ ($\leftrightarrow0.13\times1.93$\,pc$^2$ at
a distance of 7\,kpc). The second order (5.2-7.7\,$\mu$m) is observed through the
SL2 slit ($3.6''\times57''\leftrightarrow0.12\times1.93$\,pc$^2$).
The observations were done in staring mode\footnote{See the \textit{Spitzer/IRS} observer's manual at
\textit{http://ssc.spitzer.caltech.edu/documents/som/}}, consisting  in two subsequent observations centered at the 1/3 (nod 1)
and 2/3 (nod 2) positions along the slits.
When the source is being observed in a given slit (\textit{nominal observation}), the other slit, corresponding to
the other diffraction order, performs an \textit{offset observation}.
Since we are dealing with sources being more extended than the slit length, we take the
opportunity given by the offset observations to extend our measures to larger spatial
scales, $\sim 2'$.

\section{Data analysis}\label{sec:method}

\subsection{Detector image reduction}\label{sec:reduction}

\begin{figure*}
\hspace{2cm}\vspace*{1cm}\includegraphics[angle=0,width=5in,height=3.5in,clip=true]{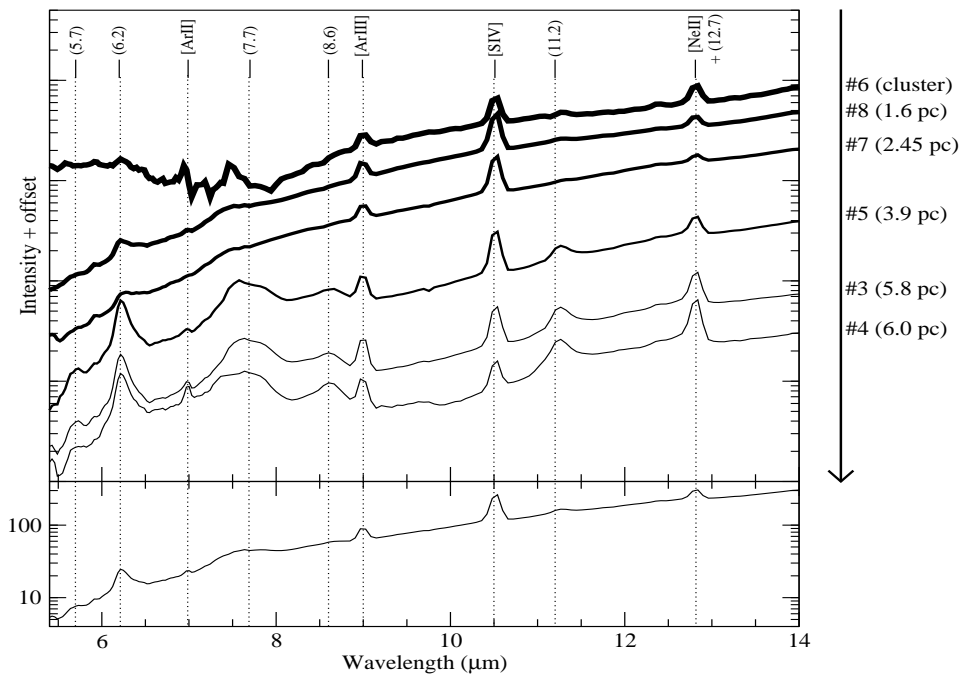}
\figcaption{In the top panel, the full slit spectra of each position (nominal observations) are sorted as a function of the projected distance from the slit center
to the central stellar cluster. In the bottom panel, we show the sum of these spectra.\label{fig:fullslit}}
\end{figure*}

The data were processed by the Spitzer Science Center, using the S13.2 pipeline.
We used the basic calibrated data (BCD) product. The background was calculated using 
the nominal and offset observations of position \#1, and substracted from the other positions.

Spectra were extracted from the two-dimensional flat-fielded detector image using scripts
within the Cornell IRS Spectroscopy Modelling Analysis and Reduction Tool environment
(SMART; Higdon et al.\ 2004). The two order spectra scale relatively well, and no
corrections were needed. The full slit spectra of the various positions and the global
spectrum of NGC\,3603 (simple sum of the spectra of all positions) are shown in
Figure\,\ref{fig:fullslit}. The final resolution is $\Delta\lambda=0.06$\,$\mu$m for SL2
and $0.12$\,$\mu$m for SL1.

\begin{figure*}\hspace{1cm}
\includegraphics[angle=0,scale=0.7,clip=true]{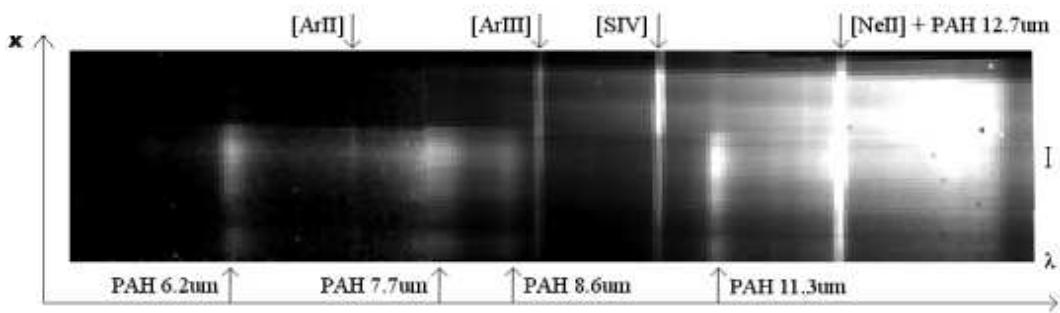}\vspace{0.1cm}
\figcaption{Spectral image of the nominal observation of position \#3. The flux (light scale) is plotted as a function of the spatial position
within the slit (y-axis) and of the wavelength (x-axis). The segment on the right side indicates the size
of the extraction windows.\label{fig:spectral_image}}
\end{figure*}

\subsection{Extraction of small-scale regions}\label{sec:smallscale}

Spatial variations can be observed within the long slits (see
Figure\,\ref{fig:spectral_image}), with regions emitting strong forbidden lines and no
PAHs, and regions showing intense PAH features together with relatively weak lines. To
investigate these spatial variations, we extracted columns, fixed to 4 pixels wide,
within the long slit on the basis that the full width at half maximum of the point spread
function (PSF) is $\approx$2 pixels. MIR sources above or below the slit, or on each side
of the extraction window within the slit, can contribute (because of the PSF) to the
total flux in a given chosen extraction window. Hence the spatial column actually
extracted corresponds to a region somewhat larger than 0.24$\times$0.12\,pc$^2$ $-$ which
would be the size of the extracted spatial region if the PSF width was null. For this
reason, it is particularly difficult to estimate the related absolute uncertainty on the flux
calibration. Given the fact that our study is focused on relative spatial variations, we
consider this caveat as being of little importance as far as the interpretations are
concerned.  Comparing the fluxes of same spatial positions observed at different locations
within the slit, we find relative errors ranging from $\pm$7\% (position \#3, lowest integrated flux)
to $\pm$4\% (position \#7, largest flux).

We extracted overlapping extraction windows along the slits, shifting by one pixel
between each other (which make a total of 37 different columns, and as many data points,
for a given nominal or offset observation). This is not an actual oversampling of the
spatial PSF. Indeed, whatever the sampling is, one is limited by the PSF convolution. The
spatial profile of spectral feature emission along the slit are smoothed by the PSF so
that the choice of the number of extracting windows only results in different
samplings of the $-$ same $-$ eventually smoothed spatial profile. Finally, it must be
added that variations of features in the observed (PSF convolved) spatial profile imply
in reality even larger variations.

\subsection{Measurements}\label{sec:measurements}

\begin{figure}[b!]\vspace*{0.2cm}
\includegraphics[angle=0,scale=.47,clip=true]{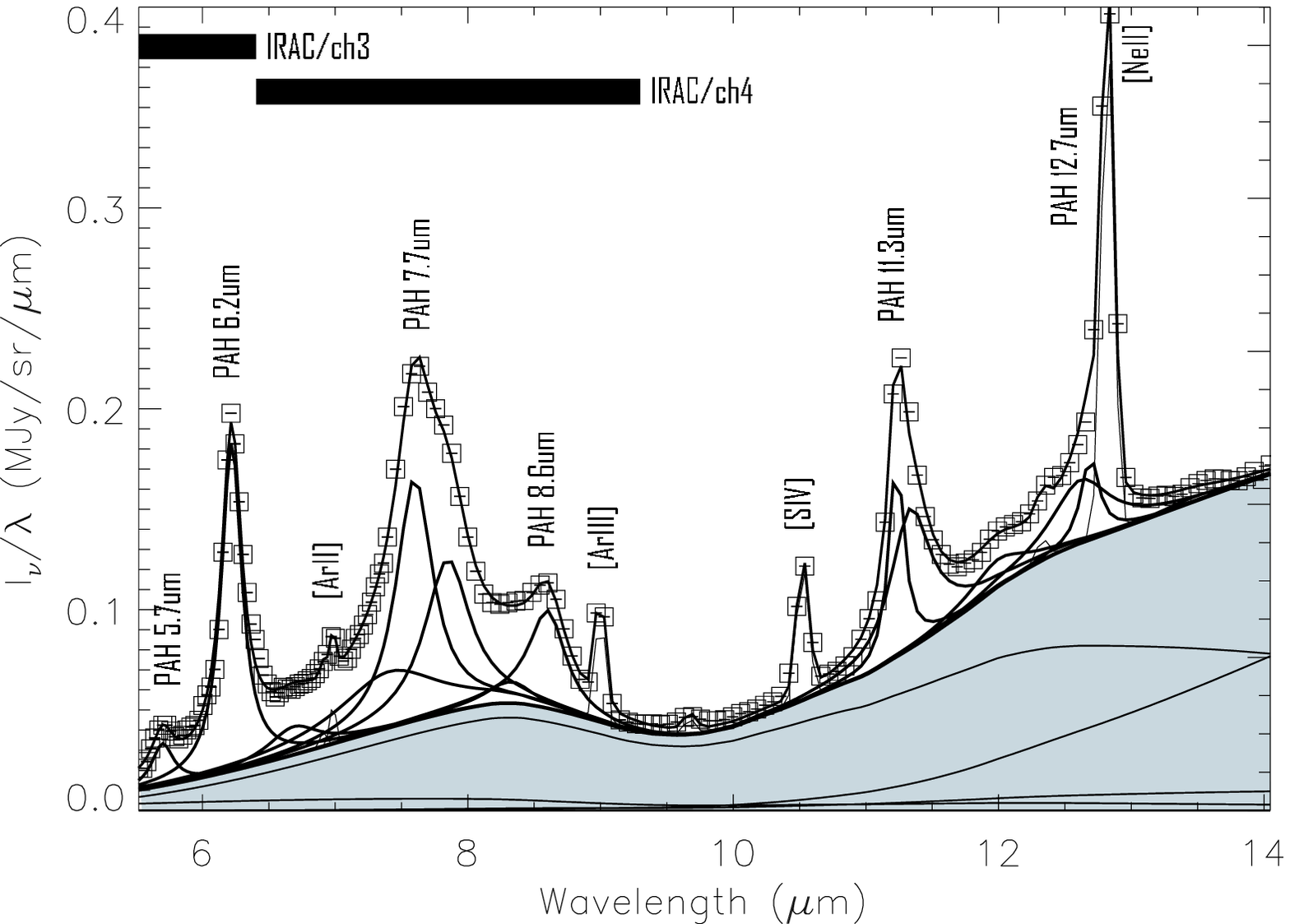}\\
\includegraphics[angle=0,scale=.47,clip=true]{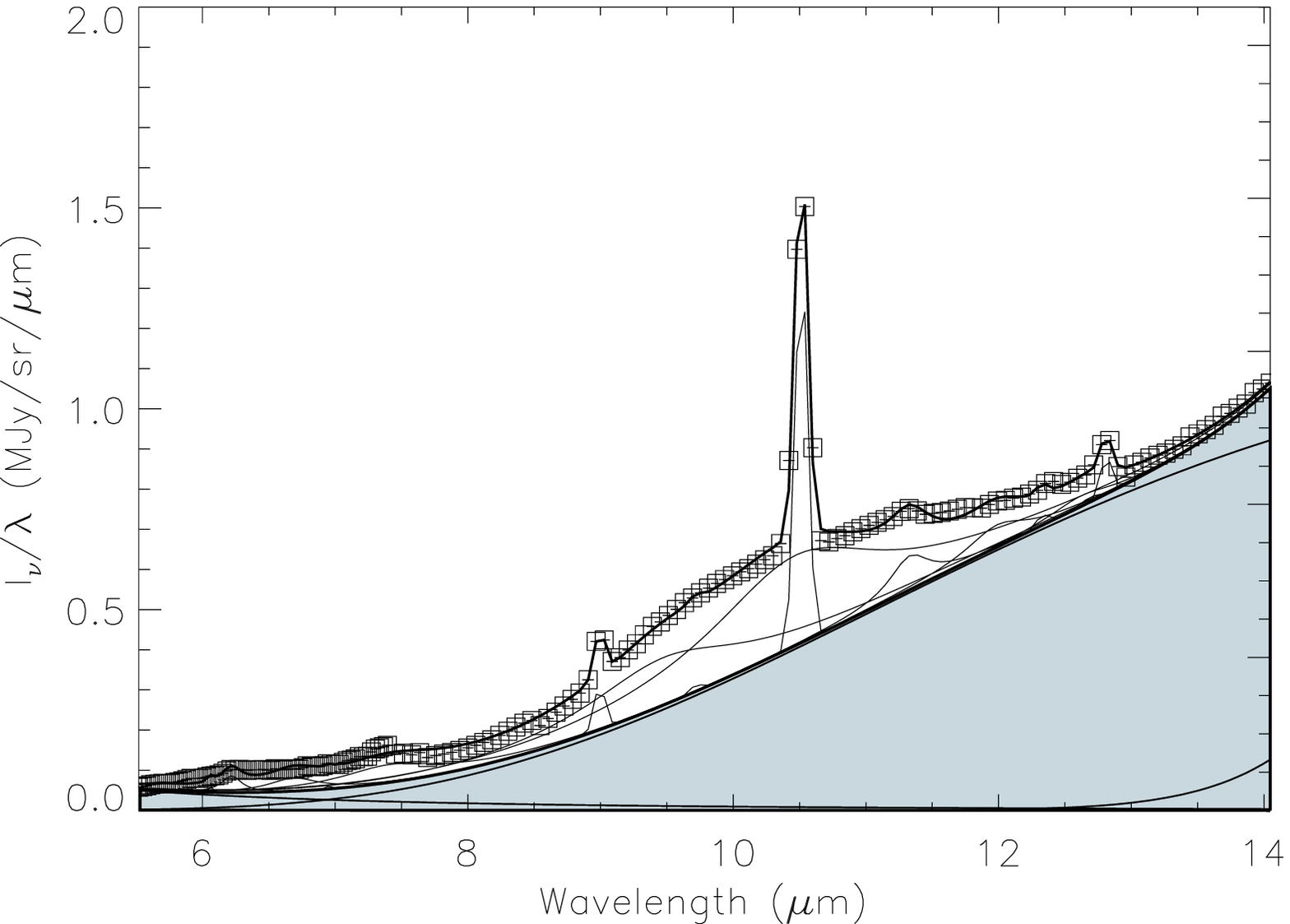}
\figcaption{Examples of the spectral fit of a 4\,pixels window in positions \#3 (top) and \#8 (bottom).
The various dust continua, PAH components, and forbidden emission lines can be identified.
We also show the coverage of the \textit{IRAC}/\textit{ch3} and \textit{ch4} bands.
The \textit{ch1} and \textit{ch2} bands observe at wavelengths shorter than 5\,$\mu$m.\label{fig:example}}
\end{figure}

We observe several PAH features and forbidden emission lines superimposed on a spectral continuum which in our case is
dominated by thermal emission of dust and broad silicate emission/absorption.
Several windows in positions \#3, \#4, \#5, and \#7 show silicate absorption around 10\,$\mu$m
while positions \#6 and \#8 show silicate emission in the same range (see Fig.\,\ref{fig:example}).
Note that the silicate features appear less prominent in the full slit spectra of Figure\,\ref{fig:fullslit} due
to the contribution of several distinct physical regions within the slit.
Only the position \#6 shows signs of a stellar emission continuum rising toward wavelengths shorter
than $\sim$8\,$\mu$m (see Fig.\,\ref{fig:fullslit}).

PAH features are detected at $\sim$5.7, $\sim$6.2, $\sim$7.7, $\sim$8.6, $\sim$11.3, and
$\sim$12.7\,$\mu$m. A weak feature at $\sim$7.9\,$\mu$m can be observed for some
positions as a shoulder of the 7.7\,$\mu$m PAH.

The following forbidden emission lines are detected, [Ne\2] (21.6\,eV) at 12.81\,$\mu$m,
[Ar\2] (15.8\,eV) at 6.99\,$\mu$m, [Ar\3] (27.6\,eV) at 8.99\,$\mu$m, and [S\4]
(34.8\,eV) at 10.51\,$\mu$m. The [Ne\2] line is blended with the PAH feature at
12.7\,$\mu$m. The [Ar\2] line could possibly be blended with a weak PAH feature at
7.0\,$\mu$m. This contamination is probably not larger than $\approx$20\% of the total
line flux (see the measurements of Sturm et al.\ 2000; F{\"o}rster Schreiber et al.\
2001).

\begin{figure*}
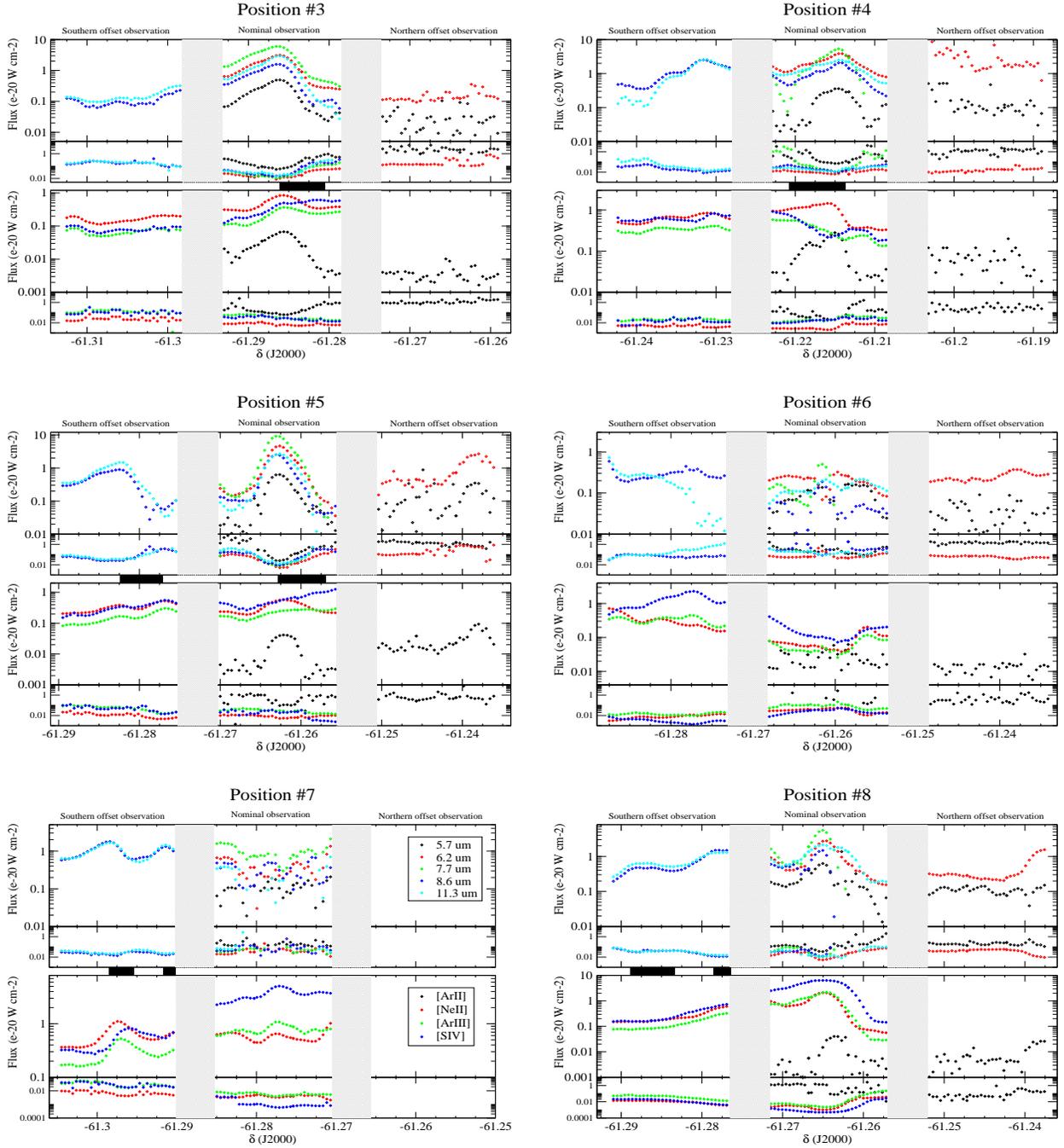
\vspace{-0.5cm}
\includegraphics[angle=0,width=3.1in,height=2.2in,clip=true]{f5a.eps}\hspace{0.5cm}
\includegraphics[angle=0,width=3.1in,height=2.2in,clip=true]{f5b.eps}\vspace{0.5cm}\newline
\includegraphics[angle=0,width=3.1in,height=2.2in,clip=true]{f5c.eps}\hspace{0.5cm}
\includegraphics[angle=0,width=3.1in,height=2.2in,clip=true]{f5d.eps}\vspace{0.5cm}\newline
\includegraphics[angle=0,width=3.1in,height=2.2in,clip=true]{f5e.eps}\hspace{0.5cm}
\includegraphics[angle=0,width=3.1in,height=2.2in,clip=true]{f5f.eps}\vspace{0.2cm}
\figcaption{Spatial variations of the spectral feature intensities. Integrated flux is
plotted against the declination for each slit position. Data on the the left side of each
plot correspond to the southern offset observation (first spectral order only,
5.2-7.7\,$\mu$m), and data on the right to the northern observation (second spectral
order only, 7.4-14.5\,$\mu$m). Legend is displayed in position \#7.  The bottom panels show the relative errors.
The black rectangles along the declination axis indicate regions where PAH intensities spatially anticorrelate
with [Ar\3] and [S\4] line intensities. The large vertical gray rectangles indicate zones
with no coverage.
\label{fig:pro3}}
\end{figure*}

The intensity measurements were done using the spectral fitting algorithm PAHFIT (Smith et al.\ 2007).
This procedure is especially suited for mid-IR low-resolution spectra of star-forming regions dominated by PAH emission.
The modelled spectral continuum is computed from a combination of starlight and 8 preset
(fixed temperature blackbodies) thermal dust continua,
whose strength is allowed to vary.
The unresolved forbidden emission lines are represented by Gaussian profiles, while the PAH features are decomposed in individual
blended Drude profiles. The main components we consider in this study are the PAH bands at
5.70\,$\mu$m, 6.22\,$\mu$m, 7.60+7.85\,$\mu$m (which we will refer from now on as the 7.7\,$\mu$m feature), 8.61\,$\mu$m, and 11.23+11.33\,$\mu$m (11.3\,$\mu$m feature).
The absorption by silicate dust, as opposed to its emission, is accounted for in the PAHFIT
calculations. We decided to model the silicate emission in positions \#6 and \#8 by including two
broad Gaussian components, centered at 9.5 and 10.5\,$\mu$m. This is an arbitrary choice
whose main purpose is to be able to measure the superimposed features with the smallest
uncertainty possible. We integrated the PAHFIT algorithm in an automatic script able to
handle the large number of spectra ($>$600) of our dataset. Examples of spectral fittings
are presented in Figure\,\ref{fig:example}. The small-scale variations of the MIR
spectral feature intensities across the slit positions are given in
Table~\ref{tab:measures} and plotted in Figure\,\ref{fig:pro3}.

Finally, in order to trace the presence of VSGs,  which are expected to dominate the
continuum emission longward of $\lambda\sim10$\,$\mu$m (D{\'e}sert et al.\ 1990), we
measure the average flux between 13.6 and 14.0\,$\mu$m,
where there is no contamination by line emission and PAH features.

\begin{deluxetable}{llllllll}
\tablecolumns{4}
\tablecaption{Flux of the lines and the PAH features.\label{tab:measures}}
\startdata
\tableline\tableline
\# & $\alpha$, $\delta$ (J2000) & d\tablenotemark{a}      &5.7\,$\mu$m & 6.2\,$\mu$m & 7.7\,$\mu$m & 8.6\,$\mu$m & 11.3\,$\mu$m  \\
   & (deg)                      &   (pc) &    &     &     &      &            \\ 
\tableline
3 & $168.70210,-61.2883 $ & $  6.000$ &$  0.2831$ &$  1.9817 $ & $ 4.2977 $ & $ 1.0332 $ & $ 1.5536 $\\ 
 &  & & (0.0516) & (0.0739) & (0.1505) & (0.0487) & (0.1001) \\ 
3 & $168.70244,-61.2880 $ & $  5.958$ &$  0.3213$ &$  2.1403 $ & $ 4.6732 $ & $ 1.2015 $ & $ 1.8042 $\\ 
 &  & & (0.0531) & (0.0677) & (0.1391) & (0.0456) & (0.0970) \\ 
3 & $168.70644,-61.2838 $ & $  5.466$ &$  0.1318$ &$  1.0722 $ & $ 1.7679 $ & $ 0.3879 $ & $ 0.8580 $\\ 
 &  & & (0.0495) & (0.0395) & (0.1166) & (0.0420) & (0.0480) \\ 
%3 & $168.70166,-61.2888 $ & $  6.055$ &$  0.2618$ &$  1.8428 $ & $ 3.9170 $ & $ 0.9376 $ & $ 1.3967 $\\ 
% &  & & (0.0401) & (0.0562) & (0.1310) & (0.0482) & (0.0769) \\ 
\multicolumn{8}{c}{\textit{(truncated)}} \\
% & ... & ... & ... & ... & ... & ... & ... \\
\tableline\tableline
\# & $\alpha$, $\delta$ (J2000) & d\tablenotemark{a}      & [Ar\2] & [Ne\2] & [Ar\3] & [S\4] & \\ % cont. @14\,$\mu$m \\
   & (deg)                      &   (pc) &    &     &     &     \\ 
\tableline
3  & $ 168.70210,-61.2883 $ & $ 6.000 $ & $ 0.0350 $ & $ 0.4403 $ & $ 0.1572 $ & $ 0.2307 $ \\ 
 & &  &  (0.0085) & (0.0103) & (0.0214) & (0.0247)  \\ 
3  & $ 168.70244,-61.2880 $ & $ 5.958 $ & $ 0.0371 $ & $ 0.5342 $ & $ 0.1861 $ & $ 0.2610 $ \\ 
 & &  &  (0.0074) & (0.0098) & (0.0155) & (0.0225)  \\ 
3  & $ 168.70644,-61.2838 $ & $ 5.466 $ & $ 0.0333 $ & $ 0.6071 $ & $ 0.2992 $ & $ 0.5256 $\\ 
 & &  &  (0.0071) & (0.0063) & (0.0150) & (0.0220)  \\ 
%3  & $ 168.70166,-61.2888 $ & $ 6.055 $ & $ 0.0271 $ & $ 0.3811 $ & $ 0.1396 $ & $ 0.2221 $ \\ 
% & &  &  (0.0067) & (0.0085) & (0.0135) & (0.0214)  \\ 
\multicolumn{8}{c}{\textit{(truncated)}} \\
% & ... & ... & ... & ... & ... & ... &  \\
\tableline
\enddata
\tablenotetext{a}{Distance from the central stellar cluster ($\alpha=11^\mathrm{h}15^\mathrm{m}07^\mathrm{s}.966$, $\delta=-61^\circ15'30''.348$).}
\tablecomments{Fluxes are expressed in $\times10^{20}$~W\,cm$^{-2}$. Numbers between brackets represent the uncertainties.}
\end{deluxetable}

\section{Interpretation of the \textit{IRAC} image}\label{sec:irac}

The \textit{IRAC} image of NGC\,3603 (Fig.\,\ref{fig:slits}) reveals a complex MIR
morphology that the spectroscopic results of the \textit{IRS} in Figure\,\ref{fig:pro3}
can help us to understand. 

We expect the \textit{IRAC/ch1} band to be dominated by stellar continuum emission and 
by the 3.3\,$\mu$m PAH feature, which is seen to scale with the 11.3\,$\mu$m feature
in various objects (Hony et al.\ 2001). As an illustration, the
full slit spectrum of the position \#6, which is centered on the central stellar cluster,
is strongly dominated by stellar emission at wavelengths shorter than 8\,$\mu$m
(Fig.\,\ref{fig:fullslit}). Furthermore, Figure\,\ref{fig:slits} shows that
stars indeed emit mostly in the \textit{ch1} band. 

The regions where PAH intensity is the largest are bright in both \textit{ch1} and \textit{ch3} bands.
This is because of the presence of PAH features (3.3\,$\mu$m and 6.2\,$\mu$m) within these bands.

The \textit{ch4} band is sensitive to the presence of the PAH features at 7.7 and
8.6\,$\mu$m, and to the forbidden emission lines [Ar\2] and [Ar\3]. By comparing the \textit{IRAC}
image with the results of the \textit{IRS}, we notice that the regions showing a flux
excess in the \textit{ch4} band as compared to the other bands are also those showing
a relatively intense [Ar\3] line in the \textit{IRS} spectra.

\section{Gas distribution}\label{sec:isrf}

The forbidden line intensities vary significantly across each observation. 
We assume on first approximation that the metallicity is uniform within the region and that it does not
affect the spatial variations we observe. Preliminary results on abundance determinations in NGC3603 
confirm the validity of this assumption (Lebouteiller et al.\ 2007, in preparation). The variations are neither due to depletion
effects, at least for argon and neon which are not expected to be
depleted onto dust grains. The depletion of sulfur is more uncertain. Abundance
determinations in H\2\ regions and planetary nebul\ae\ suggest that sulfur depletion
ranges from null to $-0.5$\,dex with respect to the solar abundance (Pottasch \&
Bernard-Salas 2006; Henry et al.\ 2004; Marigo et al.\ 2003; Mart{\'i}n-Hern{\'a}ndez et
al.\ 2002). On the other hand, the line intensity spatial variations and, to an even larger extent, the line ratios, are sensitive to the ionization
structure. Since both high-ionization lines [Ar\3] and [S\4] follow the same trend, the depletion of sulfur should not
dominate the spatial variations of [S\4] line intensity.

Because the line ratios implying species with large differences in their
ionization potentials can actually probe different regions along the sightline, we expect
\textit{a priori} the ratios [S\4]/[Ne\2], [S\4]/[Ar\3], and [Ar\3]/[Ne\2] to be our most
reliable tracers of the transversal spatial profile of the ISRF hardness.
The usual MIR estimators used in the literature are [Ne\3]/[Ne\2] and [S\4]/[S\3], which
we cannot measure with our dataset. In order to compare the reliability of our
estimators, we used the results of the stellar population and photoionization code
developed by Guiles et al. (2004). They used Starburst\,99 (Leitherer et al.\ 1999) and
MAPPINGS IIIq (Dopita et al.\ 2002; Sutherland et al.\ 2002) to derive the
variation of the MIR photoionization lines as a function of the characteristics of the
stellar population. We computed the line ratios in environments with
metallicities between 1/3 and twice solar. The stellar population in the models was
defined by an instantaneous burst, with a Salpeter initial mass function (Salpeter 1955),
and an upper mass cut-off of 100\,$M_\odot$.

The models predict that the [S\4]/[S\3] ratio traces [Ne\3]/[Ne\2], as expected, and that the [S\4]/[Ne\2],
[S\4]/[Ar\3], and [Ar\3]/[Ne\2] ratios correlate as tightly with [Ne\3]/[Ne\2]. They can
be thus considered reliable tracers of the ISRF hardness in the environments considered.
From now on, we will make use of [S\4]/[Ne\2], principally because this ratio involves
the two most prominent forbidden emission lines in our spectra.

\begin{figure}[b!]
\includegraphics[angle=0,scale=0.39]{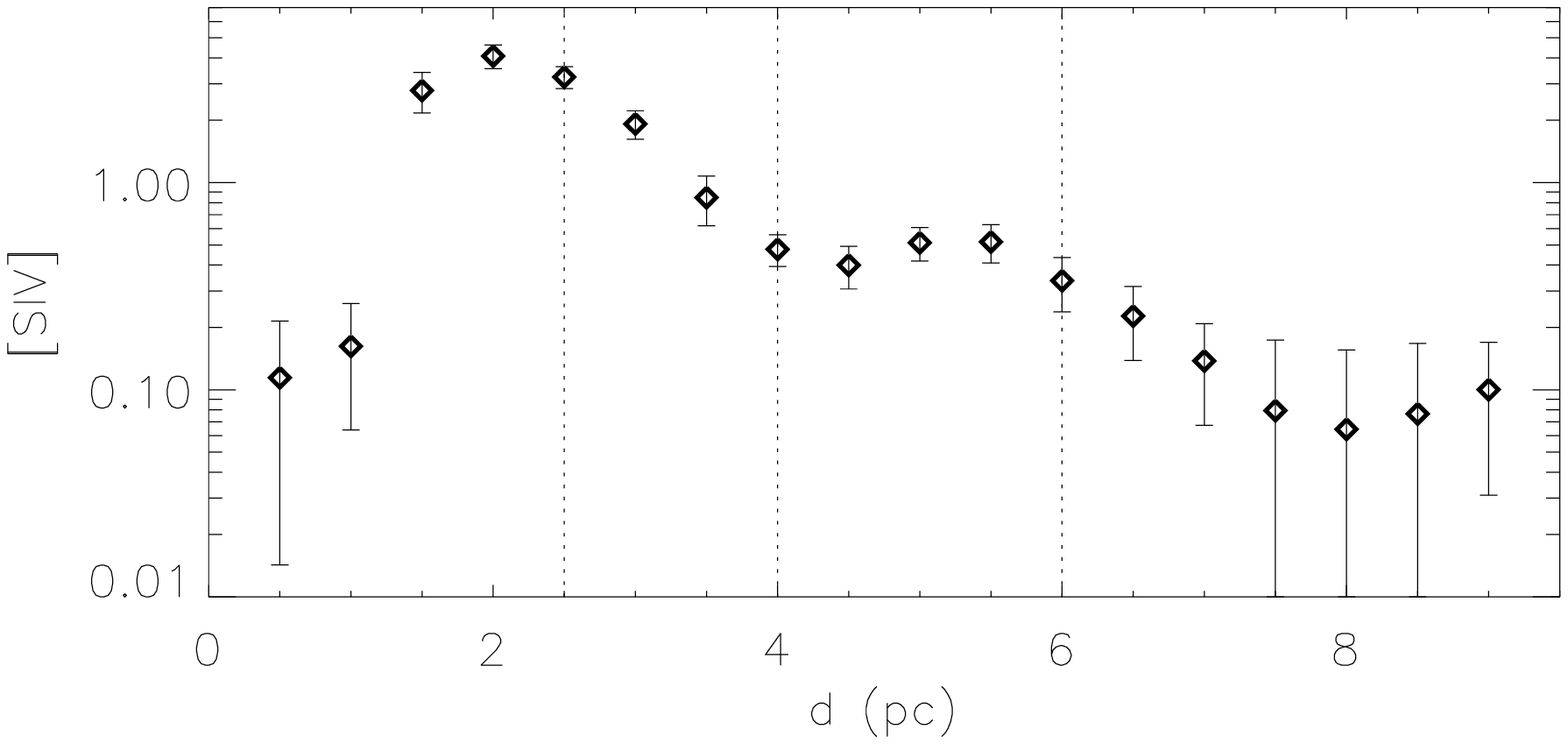}\vspace*{-0.2cm}\\
\includegraphics[angle=0,scale=0.39,]{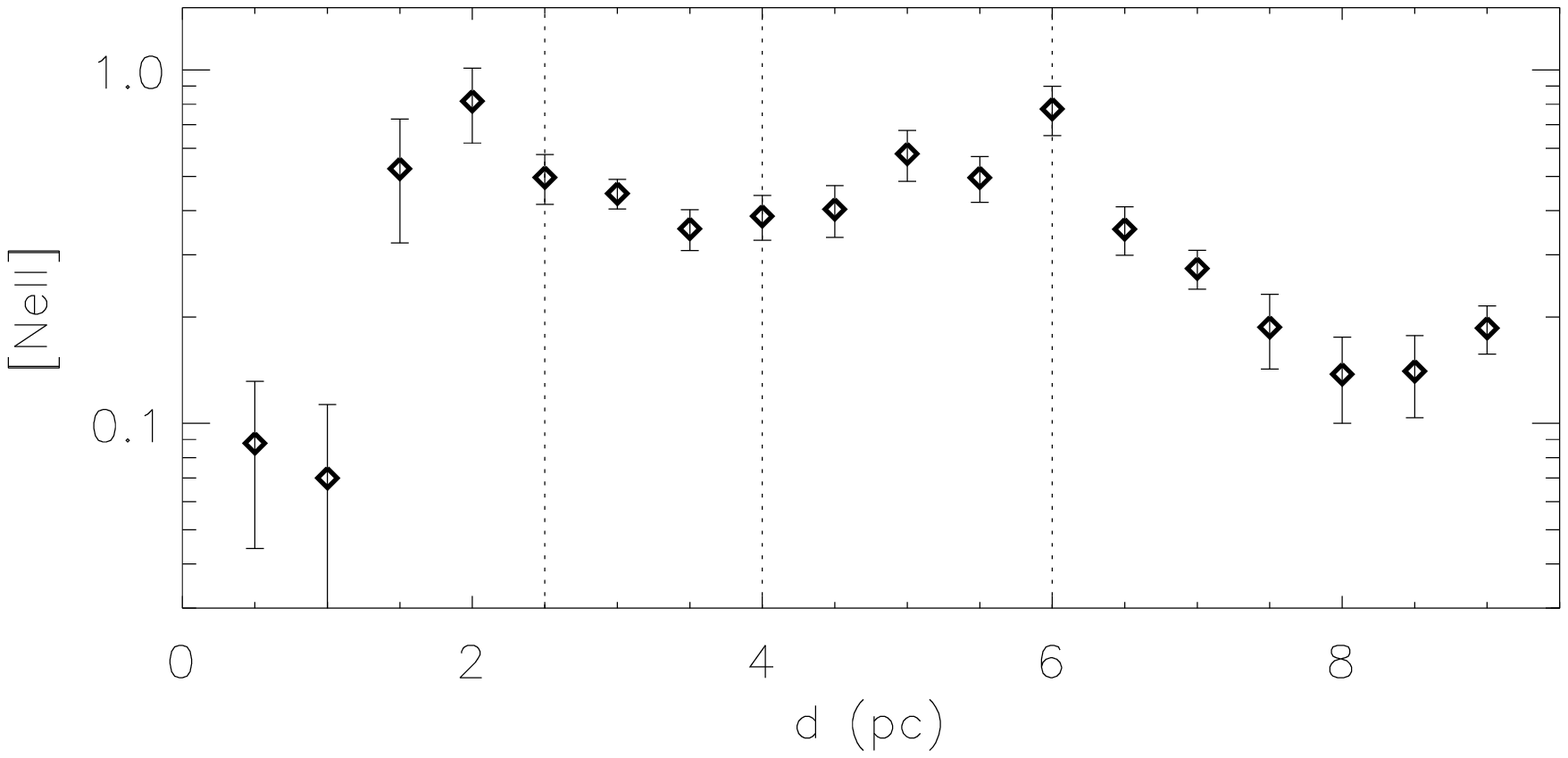}\vspace*{-0.2cm}\\
\includegraphics[angle=0,scale=0.39]{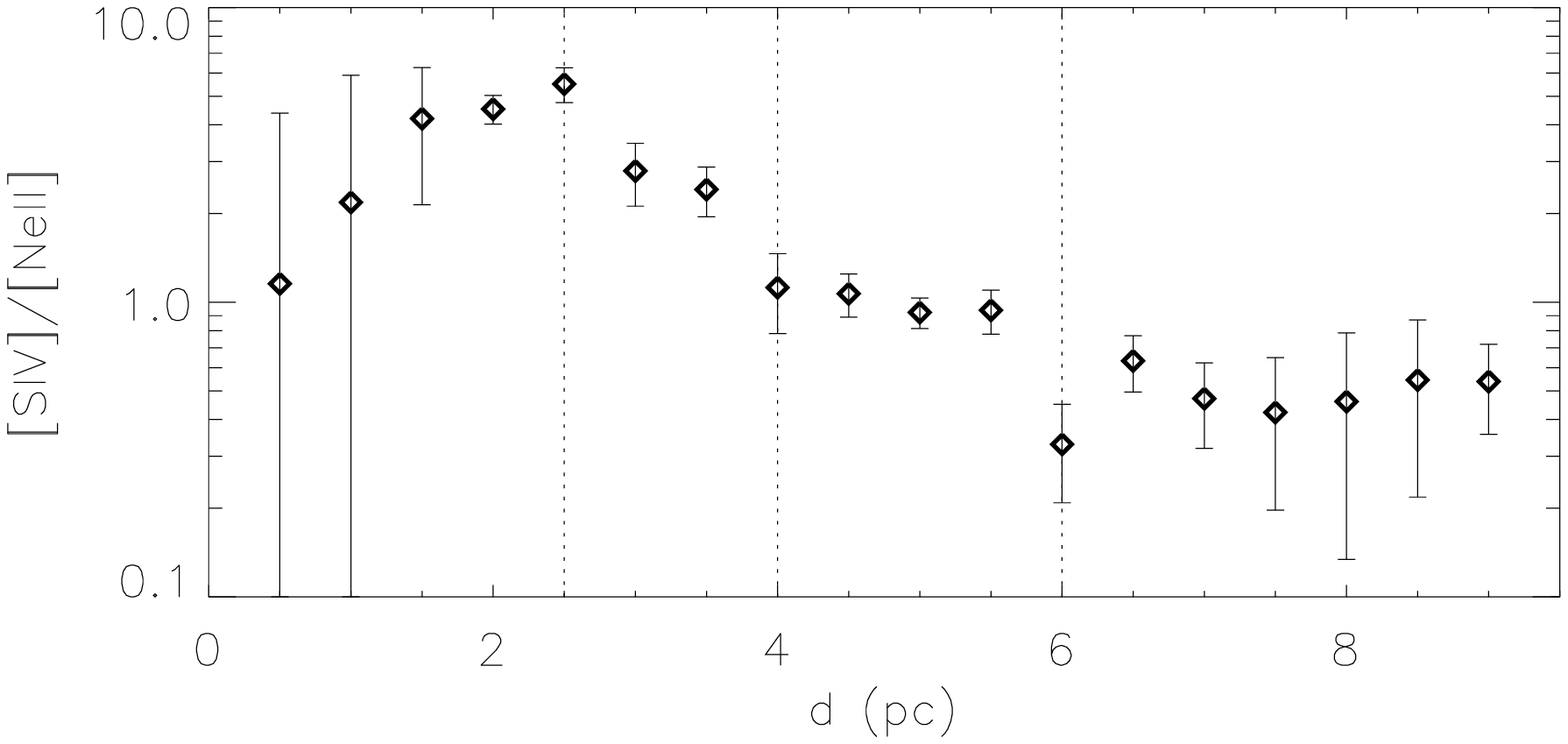}
\figcaption{[S\4] (top), [Ne\2] (middle) intensities, and the [S\4]/[Ne\2] ratio (bottom)
are plotted as a function of the projected distance from the central cluster. Intensities are expressed in $\times10^{20}$~W\,m$^{-2}$.
In order to minimize the projection effects, results in each distance bin of 0.5\,pc were error-weighted and averaged.
The vertical dotted lines refer to the distance marks plotted in Figure\,\ref{fig:slits}.
\label{fig:sivneii_dist}}
\end{figure}

We plot in Figure\,\ref{fig:sivneii_dist} the distribution of [S\4], [Ne\2], and their
ratio as a function of the projected distance from the central cluster. 
It can be seen that $I$([S\4]) shows a sharp increase between 0 and 1.5\,pc, and then decreases 
progressively until distances larger than 8\,pc. The lack of emission at very small distances from the cluster is 
probably a consequence of the stellar winds that washed out the surrounding ISM to create a cavity.  
The image of Figure\,\ref{fig:slits} shows indeed the total absence of interstellar MIR emission in the immediate vicinity
of the central cluster. Similarly to [S\4], the [Ne\2] line intensity is relatively weak at distances $\lesssim1$\,pc but 
then shows a fairly flat emission until $\sim$6\,pc.
The shallow distribution of $I$([Ne\2]) suggests that [Ne\2] emission is more extended than that of [S\4]. The [S\4]/[Ne\2] 
ratio shows roughly the same behavior as the [S\4] line intensity, and implies that the ISRF hardness decreases by a factor $\sim$20
toward the outer parts of the giant H\2\ region.

\section{Dust and molecule distribution}\label{sec:solid}

\subsection{Silicate dust}\label{sec:silicate}

\begin{figure}[b!]
\includegraphics[scale=0.41,clip=true]{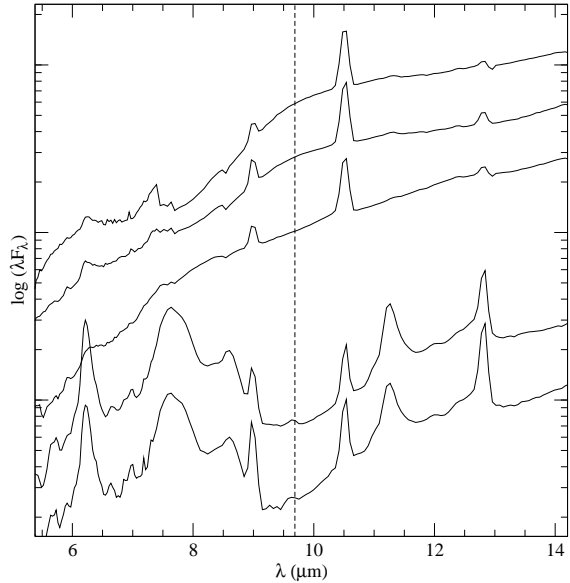}
\figcaption{The 9.7\,$\mu$m is detected in emission and in absorption. The two bottom spectra are from position \#3, the two on the
top from the position \#8, and the middle one is from position \#7.\label{fig:si_compa}}
\end{figure}

Amorphous silicate dust shows a broad spectral feature centered on 9.7\,$\mu$m, originating from the 
stretching mode of the Si$-$O bond (see, e.g., Knacke \& Thomson 1973). 
While it is mostly observed in absorption in astrophysical objects, silicate emission has also been detected 
in a few H\2\ regions, including the Orion nebula, and is thought to be due to grains with size $\gtrsim$0.01\,$\mu$m 
heated to $\sim$100\,K (Cesarsky et al.\ 2000; Contursi et al.\ 2000).
In NGC\,3603, we detect silicate dust in absorption in the spectra of positions \#3, \#4, \#5, and \#7. 
Silicate in emission is observed in the spectra of the other positions, \#6 and \#8. 
Examples of various silicate profiles across the region are presented in Figure\,\ref{fig:si_compa}.

\begin{figure}[b!]
\includegraphics[scale=0.31,clip=true]{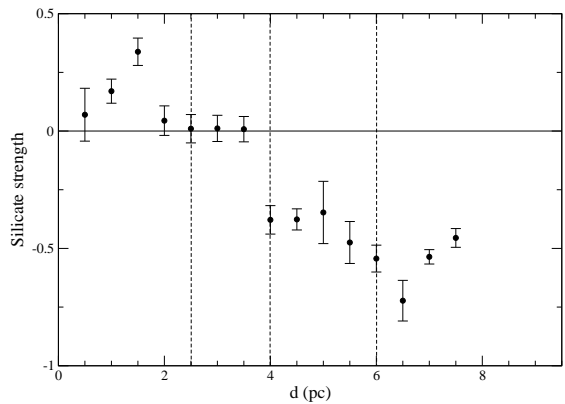}
\figcaption{Silicate strength is plotted as a function of the distance from the central
cluster. See Figure\,\ref{fig:sivneii_dist} for the plot description.\label{fig:silicate}}
\end{figure}

We measured the apparent strength of the 9.7\,$\mu$m feature using the method of Spoon et al.\ (2007).
This method requires to choose a local MIR continuum, and to calculate the ratio between the observed flux at 9.7\,$\mu$m
and the continuum value. The silicate strength, as defined by the equation (1) of Spoon et al.\ (2007), is positive for 
silicate emission, and negative for silicate absorption.
The spatial distribution of the silicate strength in NGC\,3603 is presented in
Figure\,\ref{fig:silicate}. Silicate dust is seen in emission relatively close to the
central cluster, while it is detected in absorption further away. 

It is interesting to notice that the silicate emission is observed around the same distances to the cluster as the VSG emission (\S\ref{sec:vsgs}). 
The transition zone between silicate emission and silicate absorption is
located between 2 and 3.5\,pc away from the cluster. This corresponds to the region where the ISRF hardness decreases
significantly (\S\ref{sec:isrf}) and where the PAH emission begins to dominate the spectra. 
Finally, we observe that the most prominent silicate absorption features seem to correspond to bright PAH emission regions.

\subsection{PAHs}\label{sec:pahs}

\begin{figure}[b!]
\includegraphics[scale=0.39,angle=0,clip=true]{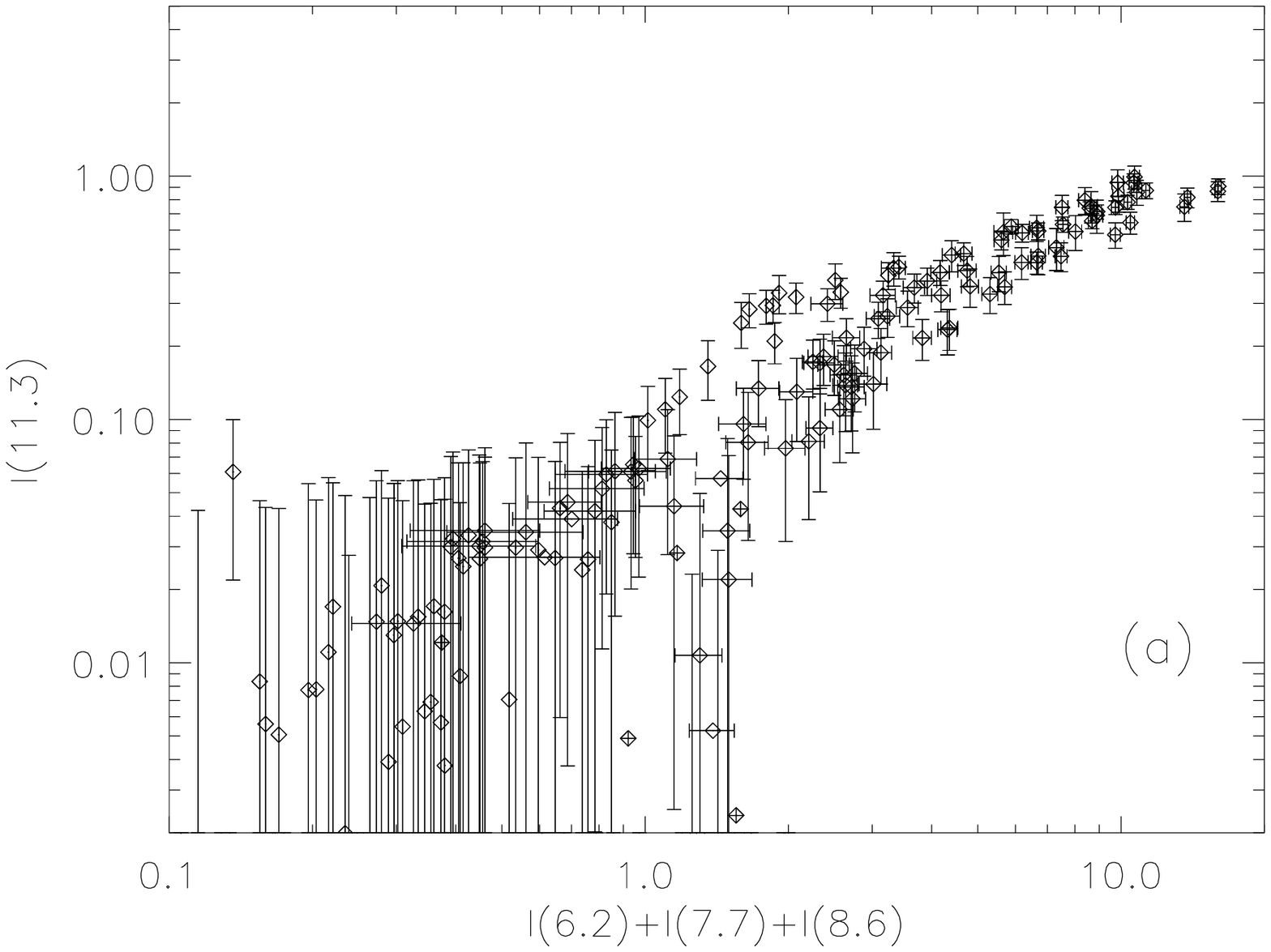}\\
\includegraphics[scale=0.39,angle=0,clip=true]{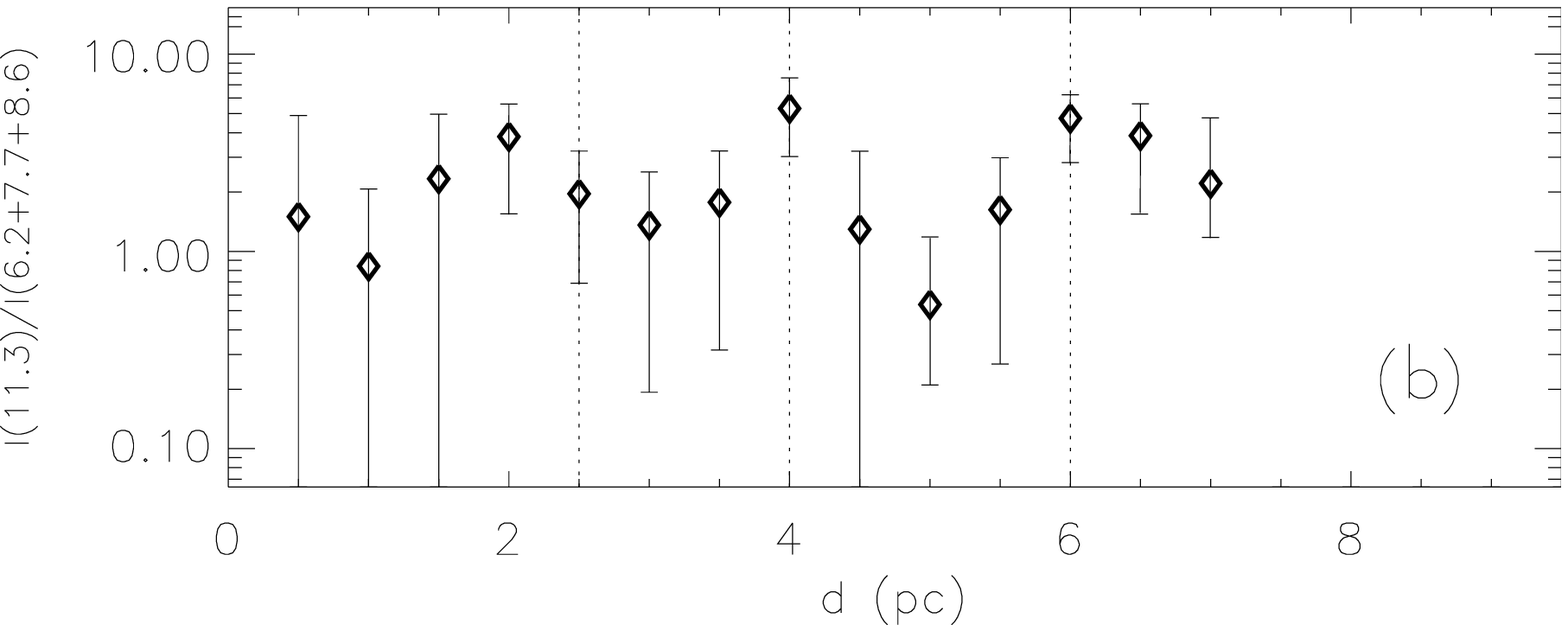}
\figcaption{(a) Intensity of the 11.3\,$\mu$m feature (neutral PAHs) is plotted against the sum of
the 6.2, 7.7, and 8.6\,$\mu$m (cations) feature intensities. (b) The ratio 11.3/(6.2+7.7+8.6) is plotted against the 
distance to the stellar cluster. See Figure\,\ref{fig:sivneii_dist} for the plot description.\label{fig:pahpah1}}
\end{figure}

%?
For the following analysis, we consider hereafter that extinction has a negligible effect on the PAH spectrum from one position
to another, or across a given observation (although it certainly has an effect on a single spectrum,
see e.g., Peeters et al.\ 2002).

The peak wavelength of the PAH features at 6.2, 7.7, and 8.6\,$\mu$m is expected to possibly shift, depending mostly on the molecule properties
(size, symmetry, ...) as opposed to the 11.3\,$\mu$m profile which is found invariably peaking at the same wavelength
(see, e.g., Hony et al.\ 2001).
However,  we find that the peak of all the PAH profiles does not vary by more
than one resolution element across the observations of NGC\,3603 (0.06\,$\mu$m for the 5.7 and 6.2\,$\mu$m features, 0.12\,$\mu$m for
the 8.6 and 11.3\,$\mu$m features), the only exception being the PAH feature at 7.7\,$\mu$m which shows a somewhat larger distribution,
centered on 7.64$\pm$0.24\,$\mu$m. %7.58 +/- 0.22 std dev
These results imply that the molecule properties do not change significantly across the region.
In particular, the invariance of the 6.2\,$\mu$m profile peak, centered on
6.23$\pm$0.06\,$\mu$m, suggests the presence of complexed and/or substituted carriers
(by introduction of an hetero atom in place of a carbon atom in the skeleton; Hudgins et al.\ 2005; Peeters et al.\ 2002). %6.23 +/- 0.03 std dev
%11.26 +/- 0.075
The spectra across NGC\,3603 identify with the "class $\mathcal A$" in the classification of Peeters et al.\ (2002), i.e., where
the PAH features actually peak at $\approx$6.22, $\approx$7.6, and $\approx$8.6\,$\mu$m. This class is populated by H\2\ regions, reflection nebul\ae\ and
most extragalactic sources.

\begin{figure}[b!]
\includegraphics[scale=0.39,angle=0,clip=true]{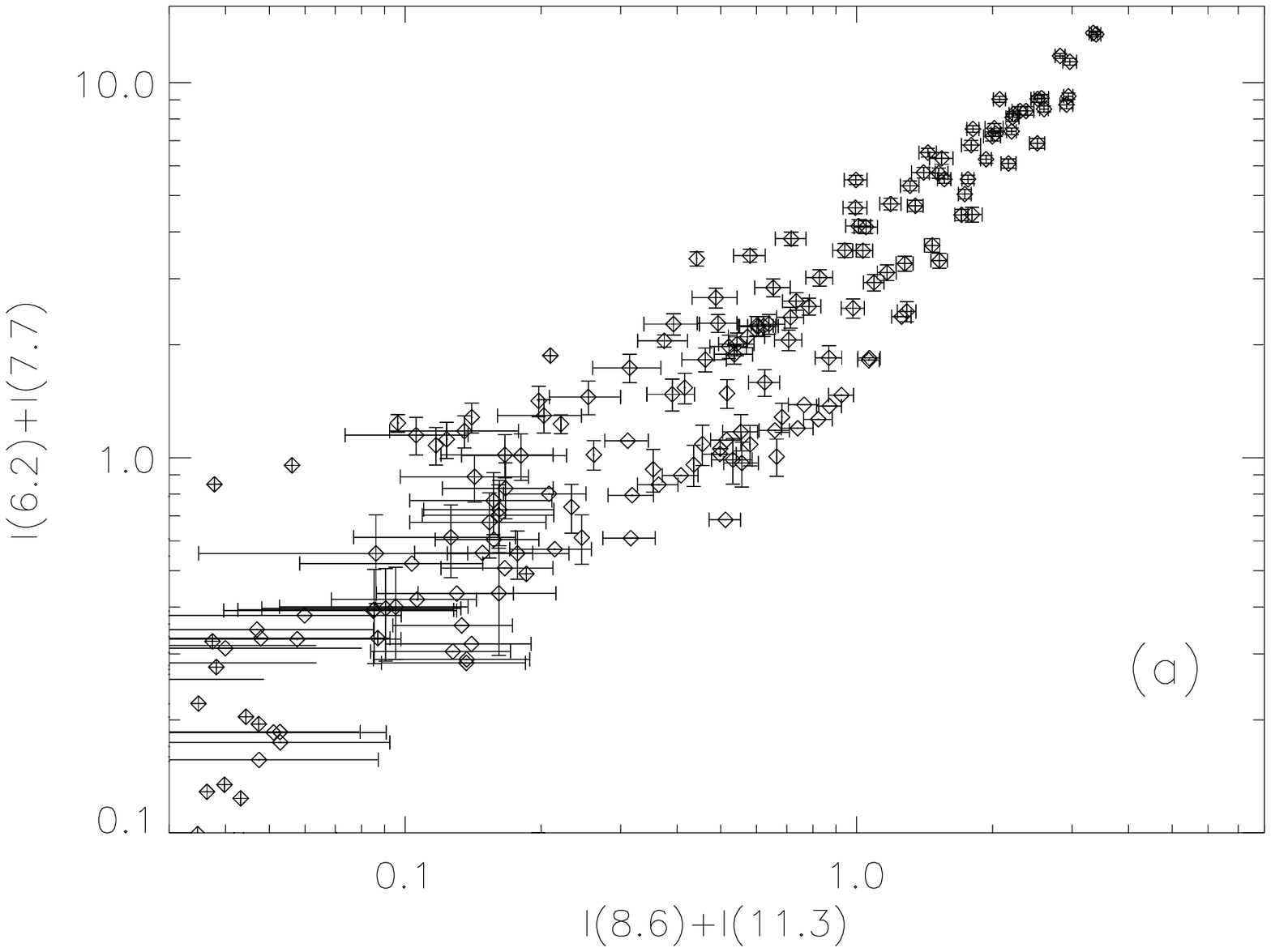}\\
\includegraphics[scale=0.39,angle=0,clip=true]{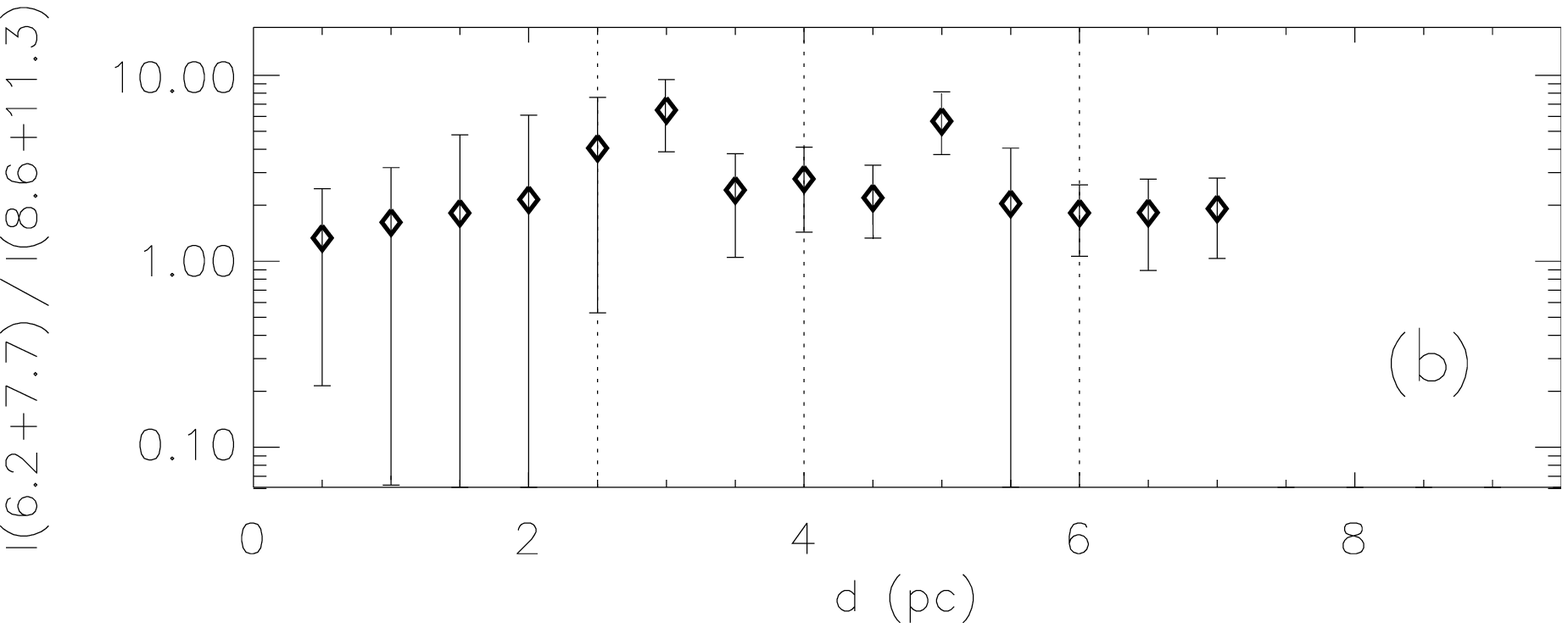}
\figcaption{(a) The summed intensity of the 6.2 and 7.7\,$\mu$m PAH features (C$-$C mode) is plotted
against the sum of the 8.6 and 11.3\,$\mu$m feature (C$-$H mode) intensities. (b) The ratio (6.2+7.7)/(8.6+11.3) is plotted against the distance to the stellar cluster.
See Figure\,\ref{fig:sivneii_dist} for the plot description.\label{fig:pahpah2}}
\end{figure}

On first approximation, the PAH spectra across the region look identical, with the
intensities of the various features scaling together. This suggests, together with the 
presence of both neutral (11.3\,$\mu$m) and ionized (6.2, 7.7, 8.6\,$\mu$m) features in the spectra, that the PAH 
ionization fraction is relatively constant. In order to investigate in more details the influence of
ionization on the PAH spectra, we compare in Figure\,\ref{fig:pahpah1}a the intensity
$I$(11.3) with $I$(6.2)+$I$(7.7)+$I$(8.6). There is a tight correlation, implying that 
the neutral/ionized mixture does not vary significantly across the region.
More particularly, the ratio is essentially constant as a function of the distance to the cluster (Fig.\,\ref{fig:pahpah1}b). 
We attribute the constant ionization fraction to the electron recombination rate in the ionized region which
balances the PAH ionization rate.

Figure\,\ref{fig:pahpah2}a shows that emission bands dominated by C$-$C modes and C$-$H modes correlate with each other. 
This correlation hold for the range of [S\4]/[Ne\2] ratios probed across the region (\S\ref{sec:isrf}).  There is no correlation with the distance to
the cluster (Fig.\,\ref{fig:pahpah2}b). These findings are consistent with the fact that the PAH ionization fraction
is constant across NGC\,3603, since ionized PAHs should show enhanced C$-$C mode emission
(see introduction).

\subsection{Very small grains}\label{sec:vsgs}

\begin{figure}[b!]
\includegraphics[scale=0.41,clip=true]{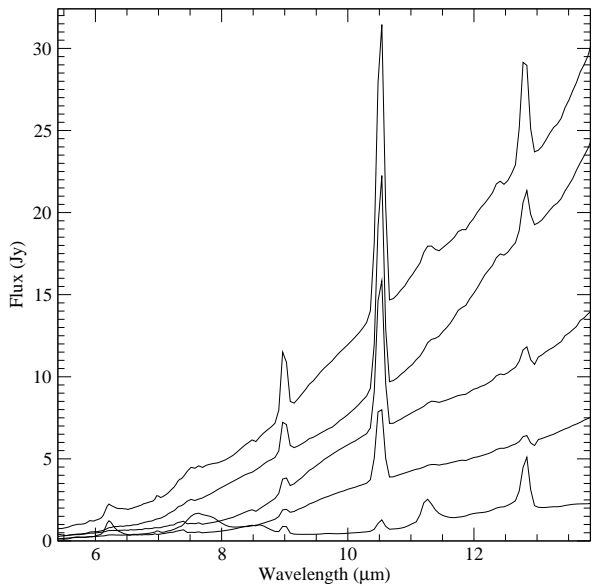}
\figcaption{Example of various spectra across NGC\,3603, showing the diversity of the VSG continuum.\label{fig:vsg_compa}}
\end{figure}

While PAH emission originates in PDRs (Tielens 1993; Sellgren et al.\ 1990), VSG
emission is seen to peak in ionized regions (Cesarsky et al.\ 1996; Verstraete et al.\
1996). In order to check whether these results hold for NGC\,3603, we estimated the VSG
emission by measuring the continuum flux at 14\,$\mu$m. The VSG emission spans a wide range
of values across the region (Fig.\,\ref{fig:vsg_compa}).

\begin{figure}[b!]
\includegraphics[angle=0,scale=0.39,clip=true]{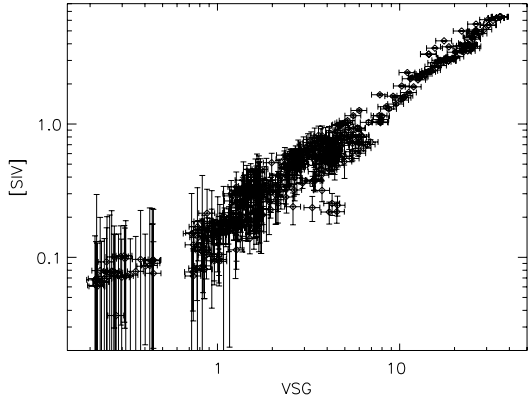}
\figcaption{$I$([S\4]), expressed in $\times10^{-20}$~W\,m$^{-2}$, is plotted against the VSG emission (in Jy).
\label{fig:nice}}
\end{figure}

We find that the VSG continuum intensity scales tightly with [S\4] line intensity (Fig.\,\ref{fig:nice}). 
Since [S\4] line intensity is seen to peak close
to the central cluster (Fig.\,\ref{fig:sivneii_dist}), we conclude that VSGs also emit mostly in these regions 
and that they are the dominant dust component in the ionized region.

\section{PAH survival}\label{sec:pahsurvival}

\subsection{Comparison with VSG emission}\label{sec:pahvsg}

\begin{figure}[b!]
\includegraphics[angle=0,scale=0.39,clip=true]{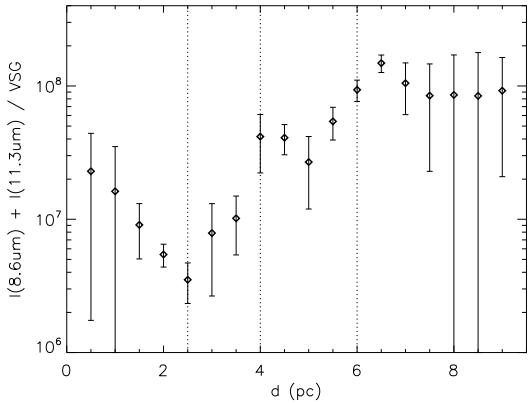}
\figcaption{The PAH/VSG ratio is plotted against the distance to the central cluster. See Figure\,\ref{fig:sivneii_dist} for the plot description.
We use the PAH features at 8.6 and 11.3\,$\mu$m only since we plot here the results of the nominal and SL1 offset observations.
\label{fig:pahvsg_dist}}
\end{figure}

The difference between the spatial emission of PAH and VSG can be seen in Figure\,\ref{fig:pahvsg_dist}, where we plot the
PAH/VSG ratio as a function of the distance to the cluster. 

\begin{figure}[b!]
\includegraphics[angle=0,scale=0.39,clip=true]{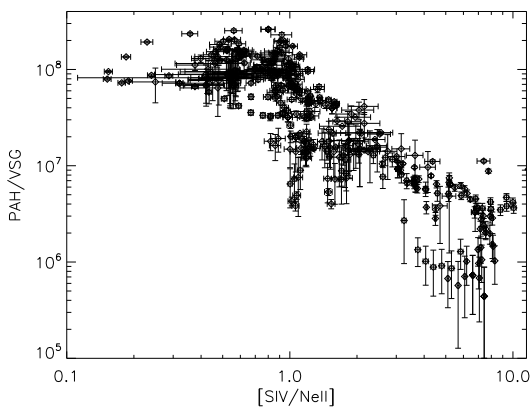}
\figcaption{The PAH intensity (sum of the 8.6, and 11.3\,$\mu$m features) is plotted against the ISRF hardness probed by the [S\4]/[Ne\2] ratio.
\label{fig:pahnom}}
\end{figure}

VSG and PAH emission do not coexist spatially, the VSGs emitting
mostly in the ionized region (although they can be present elsewhere but without being
excited). The photons exciting VSGs also illuminate PAHs, thus the PAH molecules may not
survive the physical conditions required to heat VSGs (Madden et al.\ 2006). In order to
investigate the relation between PAH emission and the hardness of the ISRF in more
details, we use a similar approach to that of Madden et al.\ (2006) by comparing the PAH
emission with VSG emission. We show in Figure\,\ref{fig:pahnom} the variation of the
PAH/VSG intensity ratio as a function of [S\4]/[Ne\2]. It can be seen that PAH emission
becomes globally weaker when the ISRF becomes harder.

The PAH features dominated by C-C or C-H emission modes do not show significant
differences across the region (\S\ref{sec:pahs}). Thus, dehydrogenation (rupture of the
C$-$H bonds by absorption of UV photons) is unlikely to be responsible for the lack of
PAHs relative to dust in regions where the ISRF is the hardest. On the other hand, PAH
molecules could be destroyed by high-energy photons.

\subsection{Presence of PDRs}

We find that [Ar\2] and [Ne\2] emission peaks always correlates with PAH emission peaks.
We find no regions where PAH emission anticorrelates with both $I$([Ar\2]) and $I$([Ne\2]).
In contrast, some regions show an anticorrelation between PAH intensity and $I$([Ar\3]) and $I$([S\4]) (see their identification in Fig.\,\ref{fig:pro3} and their corresponding locations in Fig.\,\ref{fig:slits}).
Within these regions, the location of the ion peak emission always follows a clear structure, being correlated
with the ionization potential, and implying a sharp variation of the hardness of the ISRF.

We identify these transitions as being interfaces between ionized region and PDRs. 
The interfaces toward the ionized region regime are all located in the direction the central cluster
(Fig.\,\ref{fig:slits}). The size of the interfaces (between the maximal and minimal PAH
intensities) is respectively 0.48, 0.72, 0.57, 0.66, and 0.40\,pc (note that this should
not be confused with the size of the PDR).

\subsection{Photodestruction}\label{sec:pahdestruction}

The energy deposited via photon heating is potentially large enough to dissociate bonds
within PAHs. This has been the explanation to the PAH variations in the Orion bar and in
the Red Rectangle nebula (Geballe et al.\ 1989). The Ar\2\ ion exists for energies larger
than 15.8\,eV and it is the dominant ionization stage in regions where PAH intensity is
maximal. This means that the far-UV radiation responsible for the dominant presence of
Ar\2\ (from 15.8 to 27.6\,eV which is the ionization potential of Ar\2) is not able to
destroy efficiently the PAHs, at least for the intensity of the ISRF at this energy.
Although the energy required to break C$-$H or C$-$C bonds is 4.8\,eV and $\sim$8.5\,eV
respectively (Omont 1986), photons with higher energy may be needed to dissociate bonds
in a large molecule. Using the models of Omont (1986), we find that bonds can be broken
by $\lesssim27.6$\,eV photons only in a molecule smaller than $\sim$25-50 C-atoms,
depending on the bond type. Note that the threshold size varies from one model to
another, and the values we derive should only give a first order approximation. Hence,
one possibility is that PAH molecules are large enough to prevent dissociation to occur.
Another possibility requires that PAH molecules are small, and that the energy density is
relatively weak.

Can we infer a maximal size for the PAHs in NGC\,3603? The UV radiation required for the
dominant presence of Ar\3\ (27.6-40.7\,eV) and S\4\ (34.8-47.3\,eV) ions could be
responsible for PAH molecule destruction. Single photons with energies 40.7\,eV
(47.3\,eV) are able to break bonds in molecules smaller than $\sim$40-75 ($\sim$50-85)
C-atoms, depending on the bond type. Furthermore, it also implies that the energy density
is large enough. Although even larger PAHs may be present, it is difficult to set an
upper limit on their size. However, since PAH emission is almost zero in regions where
Ar\3\ and S\4\ are dominant, it is likely that most PAHs have sizes smaller than $\sim$85
C-atoms, at least in regions where the ISRF is the hardest.

\section{MIR diagnostic diagram}\label{sec:diagnostic}

\begin{figure*}\hspace{0.5cm}
\includegraphics[angle=0,scale=.6,clip=true]{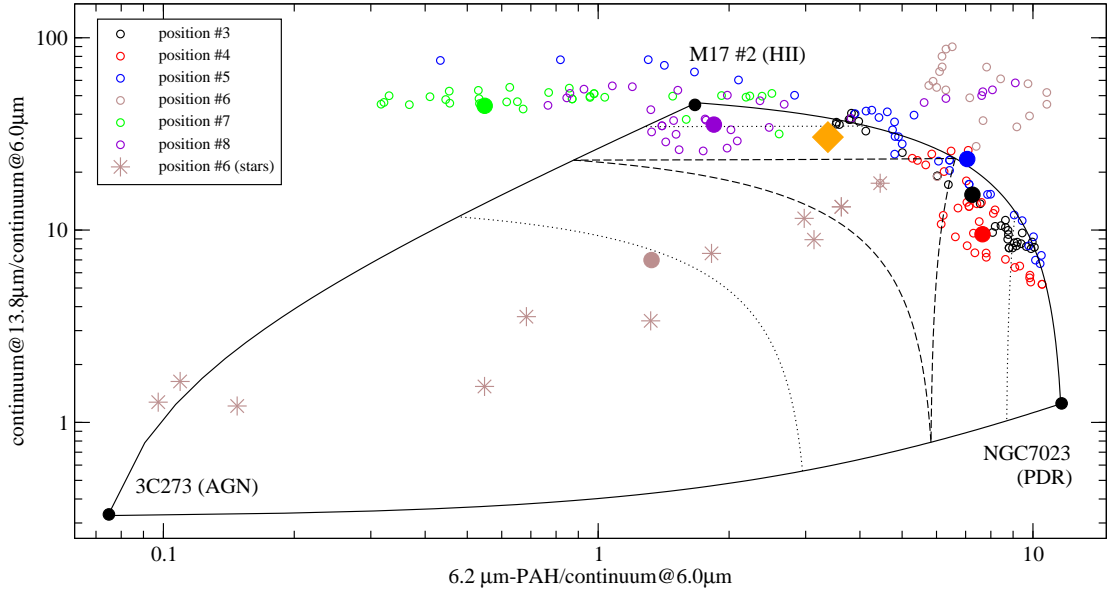}\vspace{0.2cm}
\figcaption{MIR diagnostic diagram.
The dashed and dotted curves indicate
50-50\% and 25-75\% contributions respectively. Small filled circles correspond to individual column spectra.
The stars refer to the columns within position \#6 which are centered on the central stellar cluster.
Large filled circles represent the measure using full slit extractions. The diamond stands for the measure of the global
summed spectrum of NGC\,3603.
\label{fig:laurent}}
\end{figure*}

The small-scale variations of the spectral feature intensities across NGC\,3603 show the
transition between regions dominated by PAH emission and regions dominated by
high-ionization lines. Our results should fit in MIR diagnostic diagrams used to
distinguish between active galactic nuclei (AGN), PDR, and ionized region behaviors.

In Figure\,\ref{fig:laurent}, we present a diagram, based on Laurent et al.\ (2000) and
Peeters et al.\ (2004). The templates we use are the Galactic reflection nebula NGC7023
(Werner et al.\ 2004b; PDR excited by optical photons), the H\2\ region M17 - position
\#2 (Peeters et al.\ 2004), and the quasar 3C273. Accordingly to the approach of Laurent
et al.\ (2000) and Peeters et al.\ (2004), we plot the continuum ratio between
$\sim$13.8\,$\mu$m (average flux between 13.6 and 14.0\,$\mu$m) and $\sim$6.0\,$\mu$m
(average between 5.9 and 6.1\,$\mu$m) against the ratio of the 6.2\,$\mu$m PAH intensity
over the continuum level at $\sim$6.0\,$\mu$m.

Our data points form a relatively narrow stripe, probing observationally the ionized region - PDR transition (Fig.\,\ref{fig:laurent}).
The results of positions \#7 and \#8 lie around the ionized region template, while the results of positions
\#3, \#4, and \#5 range from 100\% ionized region-contribution to 75\% PDR-contribution.
There is no region within our observed field in NGC\,3603 similar to an isolated "exposed" PDRs such as NGC7023.
This is likely due to a geometry effect because dust emission at 14\,$\mu$m could lie in the foreground or background, resulting in
an overestimation of the 14.0/6.0 continuum ratio.

The only points showing more than 50\% ``AGN-like'' behavior are those from position \#6 that are probing the central stellar cluster.
This is because for these positions, the continuum measured at 6\,$\mu$m is dominated by stellar emission. Thus, if a given starburst galaxy spectrum is
dominated by stellar cluster emission, it could \textit{a priori} be confused with an AGN when using MIR diagnostics. However, the global
spectrum of NGC\,3603, obtained using the sum of the full slit spectra of all positions (diamond in the Fig.\,\ref{fig:laurent}), does not show
any sign of stellar contamination. In this case, the regions characterized by dust emission dominate the global spectrum.
Other than the small-scale results of position \#6, none of the points probing purely ISM material show signs of AGN regime. Using the
observations in NGC\,3603, we successfully test this diagram as a diagnostic tool to distinguish regimes in a single
objects, as long as the stellar emission is not significant.

We also show in Figure\,\ref{fig:laurent} the results using full slit extraction of each
position (large squares), that were obtained by integrating the MIR emission of all the
regions within the slit. The corresponding data points of each position lie in the middle
of the small-scale results, implying that the full slit extraction does not give any
systematic error on the diagnostic.

\section{Conclusions}

We have investigated the spatial variations of the MIR spectral feature intensities (both
ionic and solid state) in the Galactic H\2\ region complex NGC\,3603, and obtained the
following results:

\begin{itemize}
\item[1.] On first approximation, the various PAH emission features have identical spatial distribution and the PAH ionization fraction
is constant.

\item[2.] The ISRF hardness, as probed by the [S\4]/[Ne\2] ratio, decreases as a
function of the distance from the central stellar cluster.

\item[3.] Silicate is detected in emission close to the cluster while it is detected in
absorption further away. Local absorption maxima of the 9.7\,$\mu$m feature seem
to identify with bright PAH emission knots.

\item[5.] PAH emission lies at larger distances than VSG emission and becomes weaker 
when the ISRF becomes harder. It has been shown
for low-metallicity galaxies by Madden et al.\ (2006).
It seems that PAH molecules are not able to survive the same physical conditions as VSGs. The simplest
explanation is the photodestruction of the molecules.

\item[6.] Small-scale results within NGC\,3603 allowed us to probe observationnally the transition between ionized
region and PDR regimes in the  MIR diagnostic diagram of Laurent et al.\ (2000). In this diagram,
the measurements using individual column spectra form a relatively narrow stripe.

\end{itemize}

\acknowledgments We thank E.\,Peeters for providing data for the diagnostic diagram
(Fig.\,\ref{fig:laurent}), S.\,Madden, G.\,Sloan, and H.\ Spoon for their useful comments. This work
is based on observations made with the {\it Spitzer Space Telescope}, which is operated
by the Jet Propulsion Laboratory, California Institute of Technology, under NASA contract
1047. Support for this work was provided by NASA through contract 1257184 issued by
JPL/Caltech.
%\email{sda@sda.org}.

%{\it Facilities:} \facility{SPITZER (IRS, IRAC)}.

\end{document}